\documentclass[letter]{aa} % for the letters 
\usepackage{natbib}
\usepackage{graphicx}
\usepackage{txfonts}
\usepackage[colorlinks=true, citecolor=blue]{hyperref}

\begin{document} 

\title{Discovery of a new nearby globular cluster with extreme kinematics located in the extension of a halo stream} 

        \author{
        Dante Minniti\inst{1,2}\thanks{To whom correspondence should be addressed; e-mail: vvvdante@gmail.com}, %\orcid{0000-0002-7064-099X},
        Jos\'e G. Fern\'andez-Trincado\inst{3}, %\orcid{0000-0003-3526-5052}, 
    Mat\'ias G\'omez\inst{1},
        Leigh C. Smith\inst{4},
    Philip W. Lucas\inst{5},
     \and 
     R. Contreras Ramos\inst{6,7}
        }
        
        \authorrunning{D. Minniti et al.} 
        
\institute{
            Depto. de Cs. F\'isicas, Facultad de Ciencias Exactas, Universidad Andr\'es Bello, Av. Fern\'andez Concha 700, Las Condes, Santiago, Chile
        \and
        Vatican Observatory, V00120 Vatican City State, Italy           
        \and
                Instituto de Astronom\'ia y Ciencias Planetarias, Universidad de Atacama, Copayapu 485, Copiap\'o, Chile
%                \and
%                {Centro de Investigaci\'on en Astronom\'ia, Universidad Bernardo O Higgins, Avenida Viel 1497, Santiago, Chile}
        \and
        Institute of Astronomy, University of Cambridge, Madingley Rd, Cambridge CB3 0HA, UK
        \and
        Centre for Astrophysics Research, University of Hertfordshire, College Lane, Hatfield, AL10 9AB, UK
                \and
       Instituto de Astrofísica, Av. Vicuña Mackenna 4860, Santiago, Chile
        \and
       Instituto Milenio de Astrofísica, Santiago, Chile
    }
        
        \date{Received ...; Accepted ...}
        \titlerunning{A New Nearby Globular Cluster Located in a Halo Stream}
        
        % \abstract{}{}{}{}{} 
        % 5 {} token are mandatory
        
        \abstract
       {We report the discovery of VVV-CL160, a new nearby globular cluster (GC) with extreme kinematics, located in the Galactic plane at $l = 10.1477$ deg, $b = 0.2999$ deg.
        }
         {We aim to characterize the physical properties of this new  GC and place it in the context of the Milky Way, exploring its possible connection with the known GC NGC 6544 and with the Hrid halo stream. }
         {VVV-CL160 was originally detected in the VISTA Variables in the V\'ia L\'actea (VVV) survey. 
         We use the proper motions (PMs) from the updated VVV Infrared Astrometric Catalog (VIRAC2)  to select GC members and make deep near-infrared color-magnitude diagrams (CMDs) to study the cluster properties. 
         We also fit King models to the decontaminated sample to determine the GC structural parameters.}
         %results heading 
        {VVV-CL160 has an unusually large PM for a Galactic GC as measured with VIRAC2 and Gaia EDR3: $\mu_{\alpha}\cos(\delta)$ = $-2.3 \pm 0.1 $ mas yr$^{-1}$ and
$\mu_{\delta}$  = $-16.8 \pm 0.1 $ mas yr$^{-1}$.  The kinematics are similar to those of the known GC {NGC~6544} and the Hrid halo stream.
We estimate a reddening of
$E(J-K)   = 1.95$ mag and an extinction of $A_{k}= 1.40$ mag for VVV-CL160. 
We also measure a distance modulus of $(m-M) = 13.01$ mag and a distance of
$D_{\odot}   = 4.0 \pm 0.5$ kpc. This places the GC at $z=29$ pc above the Galactic plane and at a galactocentric distance of $R_G=4.2$ kpc.
%halfway to the Galactic centre.
We also measure a metallicity of $[Fe/H] =  -1.4 \pm 0.2$ dex for an adopted age of $t=12$ Gyr;      
 King model fits of the PM-decontaminated sample reveal a concentrated GC, with core radius $r_{c}=   22.8"$ and  tidal radius $r_{t}=   50'$.
 We also estimate the absolute magnitude in the near-infrared of $M_{\rm k} =  -7.6$ mag, equivalent to an optical absolute magnitude of $M_{\rm v} = -5.1$ mag. 
  We also explore the possible association of this new GC with other GCs and halo streams. 
  }
        { Based on the locations and kinematics, we suggest that VVV-CL160, along with {NGC~6544}, may be associated with the extension of the Hrid halo stream. 
         %It may be speculated that the main body of this new dwarf galaxy could be located half way towards the Galactic center, hidden by extinction.
         }
%       \keywords{Galaxy : bulge -- Galaxy: globular clusters : general -- Galaxy: globular clusters: individual: NGC~6544 -- Infrared : stars -- Surveys }
        \keywords{Galaxy: bulge -- globular clusters: general -- Galaxy: globular clusters: VVV-CL160, NGC~6544 -- Surveys}
        \maketitle
        
        %%%%% INTRODUCTION %%%%%%%
        \section{Introduction}
        \label{section1}
        
Kinematically cold stellar streams are long and thin tracers of the halo \citep{Helmi1999, Combes1999, Carlberg2013}.
The recent discoveries of a number of these halo streams have helped to trace the history of mergers of infalling globular clusters (GCs) and dwarf galaxies that continue to build  the {Milky Way} \citep[MW;][]{Grillmair2017, Bonaca2018, Price-Whelan2018, Helmi2018a, Mateu2018, Carlberg2018, Malhan2018, Koposov2019, Conroy2019, Naidu2020, Ibata2020, Caldwell2020, Bonaca2021}. It is important to constrain the properties of the individual stream progenitors (GCs vs. dwarf galaxies). 

\citet{Bonaca2021} specifically predicted that GCs associated with halo  streams might be far from their parent streams (tens of degrees away).
% \textcolor{red}{such as the case of the Fimbulthul stream torn off from $\omega$ Centauri (Ibata et al. 2019)}.
Unfortunately,  halo streams cannot be traced too close to the  plane because of  severe field contamination. There is a blind region for the streams at low latitudes, and tracers such as GCs are useful for probing their total extensions. 
        
%On the other hand, 
Globular clusters  are massive ancient stellar systems that have survived destruction since the beginning of time to become fossil relics of the earliest epoch of galaxy formation \citep[e.g.,][]{Forbes2018}. Therefore, GCs can be used to characterize past accretion events that contributed to building the MW galaxy. The orbital and chemical properties of Galactic GCs have been used to identify and characterize some of these past mergers events, including the Sagittarius (Sgr) dwarf {system}, the Gaia-Enceladus-Sausage (GES), {Kraken}, and Sequoia, to list but a few \citep[e.g.,][]{Helmi2018a, Belokurov2018, Massari2019, Myeong2019, Vasiliev2019, Kruijssen2020, Forbes2020, Huang2021}.
%Barba2019, 

The discovery of new GCs toward regions of the {MW} that are heavily obscured by dust and high stellar crowding has become possible thanks to recent large near-infrared (NIR) photometric surveys such as the Two Micron All-Sky Survey \citep[2MASS;][]{Skrutskie2006} and the VISTA Variables in the V\'ia L\'actea (VVV) survey
% and the new VVV eXtended (VVVX) surveys 
\citep{Minniti2010, Saito2012}. In particular, the VVV survey
%VVVX covers $\sim$ 1700 deg$^{2}$ of the inner MW bulge and southern disk, 
is well complemented with precise infrared astrometry \citep[the VVV Infrared Astrometric Catalog, VIRAC;][]{Smith2018} and \textit{Gaia} astrometry \citep[The Gaia Collaboration;][]{Helmi2018b, Brown2020}.
%, which have yielded large numbers of GC candidates  toward the Galactic  bulge \citep{Bidin2011, Borissova2011, Borissova2014, Minniti2011, Minniti2017a, Minniti2017b, Minniti2018a, Minniti2019, Palma2019, Gran2019, Barba2019, Camargo2019, Garro2021}, disk \citep{Minniti2017b, Garro2020}, and also the Sagittarius dwarf galaxy \citep{Minniti2021}.

In this letter we analyze VVV-CL160, which we argue is a new nearby GC; it was found in the VVV survey data by \cite{Borissova2014}.
%In this Letter, we analyze a new nearby GC found in the VVV survey. Following the discovery paper of Borissova et al. (2014), the object under study here is named {VVV-CL160}. 
These authors estimated:
$E(J-Ks)=1.72 \pm 0.1$ mag,
$D=5.25 \pm 0.5$ kpc,
$[Fe/H] = -0.72 \pm 0.21$,
and age $ = 1.6 \pm 0.5$ Gyr 
for this cluster on the basis of VVV NIR photometry, concluding that VVV-CL160 was an old open cluster.
One of us later found that this is actually a GC, that we also named RCR-01 (R. Contreras Ramos 2018, private communication).
Our improvement over these previous studies is that we now count with proper motion (PM)  membership from the updated VIRAC (VIRAC2) in order to better determine the  parameters of this GC.

%The discovery of this GC accentuates the imbalance between the number of GCs located in the near-side (on our side of the Galactic center) and far-side of the MW (beyond the Galactic center).  There are many more GCs in the near-side than in the far-side, revealing that the census of Galactic GCs is still incomplete \citep{Minniti2017a}. New GCs are important not only to complete the MW census, but may also be interesting on their own. For example, the discovery of FSR~1758 \citep{Castro-Ginard2018, Barba2019} prompted the finding of the Sequoia event, a relatively massive retrograde satellite recently swallowed by the MW \citep{Myeong2019}.

Indeed, the new GC VVV-CL160 has unique characteristics that make it very interesting. It is a large cluster located halfway to the Galactic center, extending more than one degree in the sky. However, like many GCs detected by the VVV  survey, it is invisible in the optical (Fig. \ref{Figure1}). This GC has  unusual kinematics, with PMs that are very different from those of all the other Galactic GCs, with the exception of NGC~6544.
It is presently crossing the Galactic plane, located at $z\sim29$ pc above the plane. 
{We also suggest that it is located in the extension of a known halo stream. }
%, and so we suggest that it may belong to a dwarf galaxy that is being disrupted by the Milky Way.

\section{VVV near-infrared data}

%Since 2016, the VVVX survey has been operating as an extension of the completed VVV survey in order to enhance its legacy value,  providing a large spatial coverage of the inner Galaxy from Galactic longitude between 20 deg ($l<20$ deg) to $-130$ deg (or $l>230$ deg). 
The VVV survey data were acquired with the 4.1 m  Visual and Infrared Survey Telescope for Astronomy \citep[VISTA;][]{Emerson2010} located at the ESO Paranal Observatory.\ It is equipped with the 16-detector VISTA Infrared  CAMera \citep[VIRCAM;][]{Dalton2006}, with a field of view of $1.1\times1.5$ deg$^{2}$, and operates in the 0.9--2.5$\mu$m wavelength range.  

The data were reduced with the VISTA Data Flow System \citep{Emerson2004} at the Cambridge Astronomical Survey Unit \citep[CASU;][]{Irwin2004}. Processed images and photometric catalogs are available from the ESO Science Archive and from the VISTA Science Archive \citep{Cross2012}.

%We followed the same procedure as described in detail by Smith et al. (2018) to extract our own point spread function (PSF) photometry in the region around {VVV-CL160}, in order to obtain the best result of this crowded environment.
%  (VIRAC2, Smith et al. 2021, in preparation). 
%The photometry of saturated stars (usually with $K_{s} \lesssim$ 11.5 mag) was also obtained with VIRAC2 from the wings of the PSF. 
We used preliminary data from VIRAC2 (Smith et al. 2021 in preparation), which features a number of enhancements over the original VIRAC \citep{Smith2018}. These include  point spread function (PSF)  photometry and astrometry, the use of Gaia Data Release 2 (DR2) as an astrometric reference frame, a global photometric calibration, and other pipeline enhancements. Point spread function fitting in particular is necessary to obtain robust results in the crowded environment of VVV-CL160. We note that the new Gaia EDR3 PMs for this cluster are consistent with the VIRAC2 measurements.

\section{Confirmation of a new globular cluster}

%{VVV-CL160} was identified as a previously unknown stellar cluster by 

\begin{figure}
        \begin{center}
                \includegraphics[width=90mm]{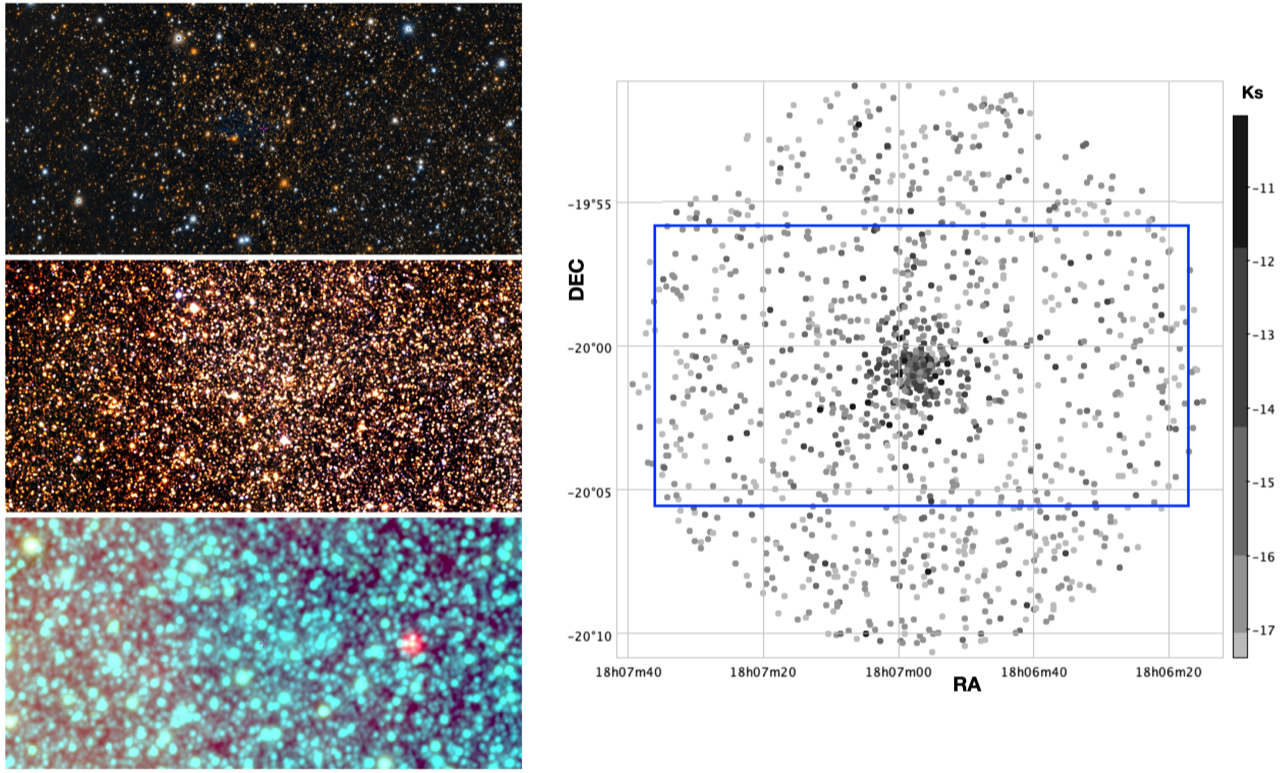}
                \caption{Finding charts for VVV-CL160 at different wavelengths.
                Left panels: 
                 Field covering $10' \times 20'$ centered on VVV-CL160 in the optical from Pan-STARRS (top), NIR from 2MASS (middle), and mid-infrared from WISE (bottom). 
                 Right panel: Map showing only the PM-selected sources, with the blue rectangle showing the same field as the left panels.
                }
                \label{Figure1}
        \end{center}
\end{figure}

Borissova et al. (2014) classified VVV-CL160 as an old ($t=1.6$ Gyr) and metal-poor ($[Fe/H]=-0.72$) open cluster. 
%located in the Galactic plane at $GL = 10.15$ deg, $b = 0.30$ deg. 
This cluster is located in a very reddened and crowded region of the Galactic plane at $l = 10.15$ deg, $b = 0.30$ deg; as a consequence, it is invisible in optical images (see Fig. \ref{Figure1}). Unfortunately, this means that there are very few cluster sources that can be measured by Gaia. \citet{Borissova2014} also obtained radial velocities for two red giant stars, with $RV_1 = 185 \pm 10$ km/s and $RV_2 = 20\pm 14$ km/s, arguing that only  object 1 is a cluster member,  object 2 being a field star. 
Both objects are saturated in the VVV images, although they have been measured in Gaia EDR3.
Object 1 has Gaia $ID=4094961256139633920$, 
%located at   $RA=271.74042115$,  $DEC=-20.00762145$ 
with $PMRA = -4.335 \pm 0.242$ mas/yr,  $PMDec = -6.514 \pm 0.189$ mas/yr.
Object 2 has Gaia $ID=4094960878182514048$, 
%at  $RA=271.73614001$  $DEC=-20.00951485$, 
with $PMRA = 0.345 \pm 0.020$ mas/yr,  $PMDec = -0.846 \pm 0.015$ mas/yr.
Given these Gaia EDR3 PMs, neither of these red giants appears to be a cluster member.
% according to the Gaia EDR3 PMs.

Because the interstellar medium is more transparent in the NIR, we see not only the foreground disk stars but also the background bulge and disk populations. This is another instance where the field contamination largely outnumbers the cluster member stars. 
%This cluster therefore is barely seen even in the near-IR images. 
The {VIRAC2} NIR PMs (Smith et al. 2021, in preparation) are crucial for establishing GC membership, which in turn is needed to make clean color-magnitude diagrams (CMDs) that allow the measurement of the physical cluster parameters.

\begin{figure}
\begin{center}
                \includegraphics[width=84mm]{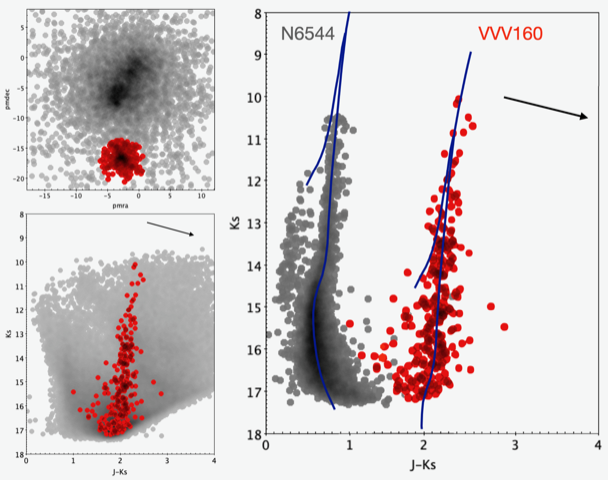}
                \caption{Decontaminated CMDs for VVV-CL160. Top left: VIRAC2 PMs in the GC region, with field stars in gray and selected GC members in red. Bottom left: VIRAC2 NIR CMD for the same region. Right: Decontaminated VVV-CL160 CMD compared with the known Galactic GC {NGC~6544}. The fitted $\alpha$-enhanced isochrones with $[Fe/H] =  -1.4 $ dex and an age of $t=12$ Gyr are shifted to their respective reddenings and distances (see text). The direction of the reddening vector is shown with the arrow.
                }
                \label{Figure2}
        \end{center}
\end{figure}

The top left panel of Fig. \ref{Figure2} shows the PM selection from the VIRAC2 data in this field, with the cluster members indicated in red, on top of the background bulge and disk population, shown in gray. 
The bottom left panel shows the cluster CMD (in red) compared with the overwhelming number of field stars (in gray), which outnumber the GC stars by a factor of about 100. The field stars from the Galactic disk and bulge along this line of sight are located at different distances and suffer different amounts of reddening and extinction, and the cluster red giant branch (RGB) is clearly seen after PM selection. The right panel of Fig. \ref{Figure2} finally shows the CMD of VVV-CL160 compared with that of the well-studied GC NGC~6544. They are both made with the VIRAC2 PSF photometry; we chose NGC~6544 in particular because it is located in the same region of the sky and it has similar metallicity and mean motions. The VVV-CL160 shows not only a tight RGB, but also an excess of faint blue sources corresponding to the blue horizontal branch (BHB) as well as {an} asymptotic giant branch (AGB) bump at $Ks =13.5 \pm 0.1$ mag.

We also explored the Gaia EDR3 data in this region \citep{Brown2020}, finding that only a dozen stars have PMs consistent with VVV-CL160. These are the brightest cluster giants, which are, however, very faint and red in the Gaia CMDs. Their mean PMs are in agreement with the VIRAC2 measurements, and their parallaxes are too uncertain to determine a reliable geometric cluster distance.

In order to determine the distance to VVV-CL160, we made optical and NIR CMDs via comparison with the well-studied nearby Galactic GC {NGC~6544}, 
{located at a distance of $D=2.6$ kpc with a distance modulus of m-M=12.07 mag  \citep{Vasiliev2021}. 
}
Both clusters show similar RGB and BHB morphologies, and their sequences are separated by $\Delta(J-Ks)=1.30$ mag, with VVV-CL160 being more reddened, and $\Delta Ks= 2.40$ mag, with VVV-CL160 being fainter. Taking into account the fact that the NGC 6544 reddening is $E(J-Ks)= 0.40$ mag \citep{ContrerasRamos2017}, this implies that the VVV-CL160 reddening is $E(J-Ks)= 1.70$ mag, similar to the value {obtained} by \citet{Borissova2014}. The differential comparison with NGC 6544 (Fig. \ref{Figure2}) yields a distance modulus of $m-M = 13.01 \pm 0.10$ mag, equivalent to $D=4.0 \pm 0.2$ kpc.

We measured $E(J-K)=1.95$ mag and $Ak = 1.40$ mag from the reddening maps of \citet{Surot2020}, which is slightly larger than the value obtained from the comparison with NGC 6544. A difference in reddening of $0.1-0.2$ mag would not be surprising because the reddening maps are integrated along the line of sight to the bulge and VVV-CL160 is likely placed in front of the bulge, slightly less reddened. In addition, within the cluster field there is differential reddening observed in the maps of \citet{Surot2020}, ranging from about $E(J-Ks) = 1.6$ mag to about $2.0$ mag.

We  measure an integrated apparent magnitude of $Ks=6.84$ mag, which at the distance of $D=4$ kpc  yields a total absolute magnitude of $M_K=-7.6$ mag. This is equivalent to $M_V = -5.1$ mag, making VVV-CL160 a new low-luminosity Galactic GC.
%
%Adopting the reddening values from the maps of \citet{Schlafly2011} we obtain $E(J-K)=4.06$ mag and $A_K = 3.01$ mag, which would imply a larger distance modulus m-M=14.62, yielding a total absolute magnitude $M_K=-9.2$ mag. This is equivalent to $M_V=-6.7$, which would place its luminosity closer to the peak of the Galactic GC luminosity function.

        \begin{figure}
        \begin{center}
                \includegraphics[width=62mm]{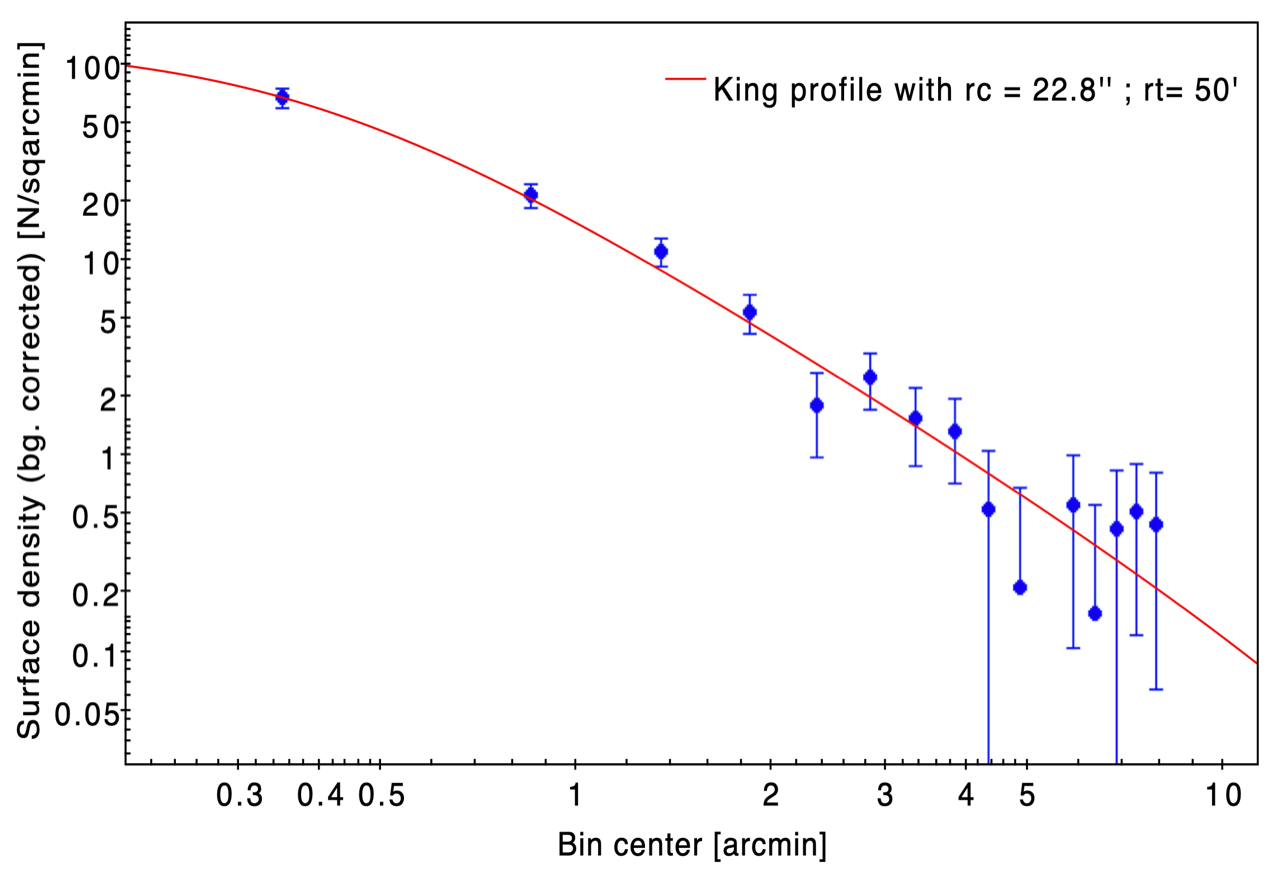}
                \caption{Radial dependence of the PM-decontaminated surface density fitted using a King model with the indicated radial parameters.}
                \label{Figure3}
        \end{center}
\end{figure}

We derive the location of the GC center at: $RA_0 = 271.738 \pm 0.01$ deg,  $Dec_0 = -20.015 \pm 0.01$ deg, and we fitted a King profile to the PM-selected sources, as shown in Fig. \ref{Figure3}. The core radius is well defined, revealing a concentrated cluster with $r_c = 22.8”$, equivalent to 0.44 pc at the distance of 4 kpc. Given the high background, the tidal radius is less well defined, and after different trials we settled on a value of $r_t=50’$, equivalent to 58 pc at that distance. A fit with $r_t=1$ deg is also as good, and we cannot exclude the possibilities that this cluster is more extended or that it possesses extra tidal stars. In any case, this appears to be a very concentrated GC with $log(r_c/r_t) = -2.1$.

        \begin{figure}
        \begin{center}
                \includegraphics[width=52mm]{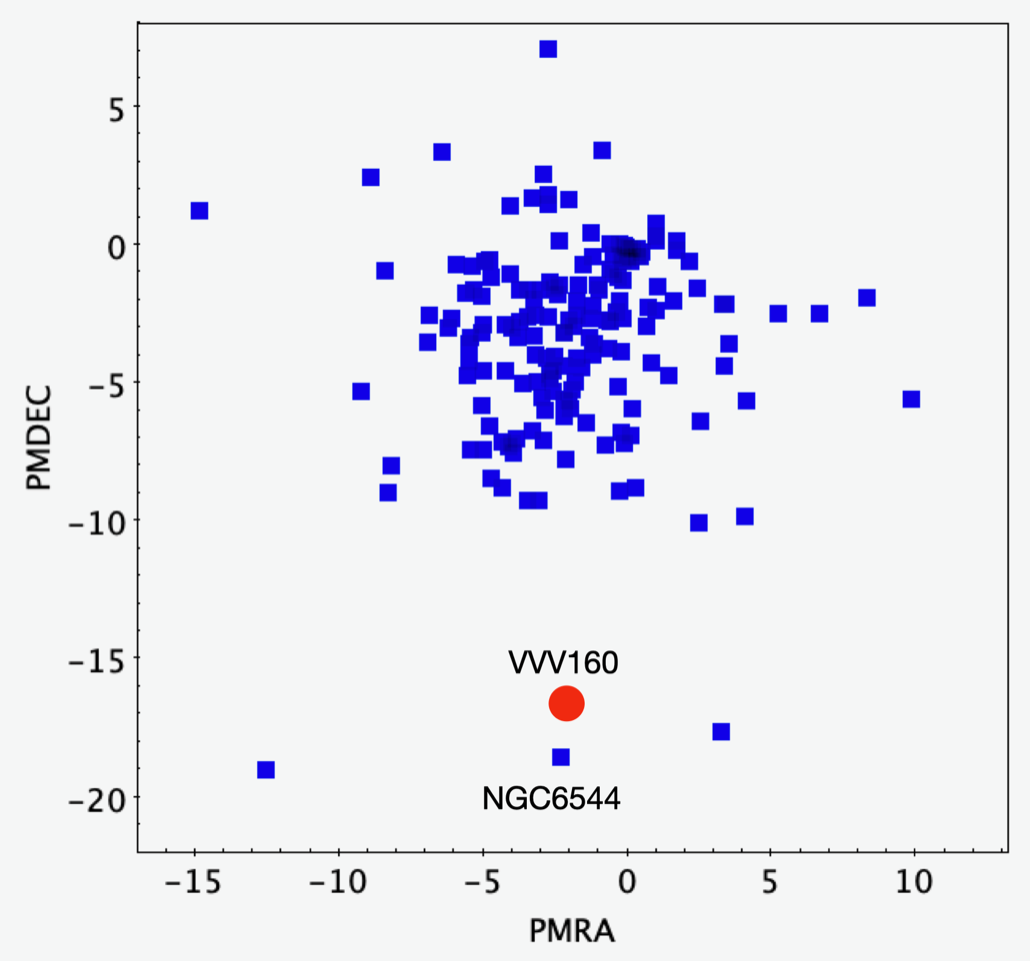}
                \caption{Vector PM diagram showing the mean PMs in units of mas/yr for all Galactic GCs from Vasiliev \& Baumgardt (2020), in blue, compared with the mean PM of VVV-CL160, in red. This illustrates that the only other GC that shares similar kinematics is NGC 6544.}
                \label{Figure4}
        \end{center}
\end{figure}

We measure the following mean PMs with VIRAC2 and Gaia EDR3: $\mu_{\alpha}\cos(\delta)$ = $-2.3 \pm 0.1 $ mas yr$^{-1}$ and
$\mu_{\delta}$  = $-16.8 \pm 0.1 $ mas yr$^{-1}$. Figure \ref{Figure4} shows the mean PM of VVV-CL160 (red circle) compared with the mean PMs for all the Galactic GCs from \citet{Vasiliev2021}. It is evident that this cluster is kinematically very different from the rest of the GCs in the MW, except for NGC 6544, the only GC to share a similar PM. The tangential motion of VVV-CL160, directed almost perpendicular to the Galactic plane, is also unlike any other of the MW populations expected along the line of sight as it is inconsistent with kinematic models of the Galactic bulge field stars \citep[e.g.,][]{Brunetti2010}. In fact, this extreme behavior is only similar to that of the GC NGC 6544 and the Hrid halo stream, as discussed in the next section.

\begin{figure}
        \begin{center}
                \includegraphics[width=74mm]{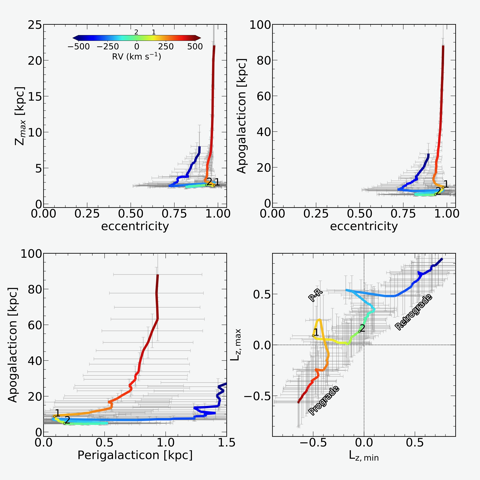}
                \caption{Predicted orbital motion for VVV-CL160 colored according to a wide range of expected radial velocities, from $-500$ km/s (blue, retrograde orbits), through 0 km/s (green), to 500 km/s (red, prograde orbits). Top left: Maximum height above the Galactic plane Z-max in kpc vs. orbital eccentricity.
                Top right: Apogalactic distance in kpc vs. orbital eccentricity.
                Bottom left: Apogalactic distance in kpc vs. perigalactic distance in kpc. Bottom right: {Maximum and minimum z-component of the angular momentum}.
                We note that the location of the two red giant stars with measured radial velocities from Borissova et al. (2014), which are labeled 1 and 2, have $RV_1 = 185 \pm 10$ km/s and $RV_2 = 20\pm 14$ km/s, respectively.
                }
                \label{Figure5}
        \end{center}
\end{figure}

Figure \ref{Figure5} shows the {predicted} orbital {elements} for VVV-CL160, computed using the {\texttt{GravPot16} model}\footnote{https://gravpot.utinam.cnrs.fr} 
\citep[see][]{Fernandez-Trincado2020, Fernandez-Trincado2021} and assuming different radial velocities in the range $-500 < RV< 500$ km/s. The orbits corresponding to the radial velocities of the two objects, 1 and 2, from \citet{Borissova2014} are labeled. These are non-members according to the Gaia EDR3 PMs, but we use them to illustrate two typical radial velocities.
On the basis of this figure, we suggest that, for reasonable radial velocity values, this GC might have an eccentric orbit with $Z_{\rm max} \sim 2.5$ kpc, $R_{\rm peri} \sim 0.1-0.3$ kpc, and $R_{\rm apo} \sim 5-10$ kpc, {changing the sense of motion from prograde to retrograde (P-R) at the same time; this could be due to the presence of the Galactic bar}. 
However, accurate radial velocities are needed.
% for VVV-CL160.
%Also, if NGC 6544 and VVV-CL160 are associated with the Hrid stellar stream, this will help to better constrain the orbital path and history of the parent dwarf galaxy.

%\section{Summary and discussion}

%\section{Potentially associated Planetary Nebulae}

We note that {VVV-CL160} is also interesting because % it contains a PN and XRS within 3’ of its centre.
there are three candidate planetary nebulae (PNe) in the vicinity: The closest one is PHR J1806-1956 at 4.6’ north of the GC center; then there are PN PM 1-209 and IRAS 18036-1954 at 4.9’ west and 5.2’ northwest from the center, respectively. Their membership in the GC is unknown; determining this membership spectroscopically would be significant as there are a very limited number of GC PNe known in our Galaxy \citep{Minniti2019}.
Of these three PNe, as far as we know only IRAS18036-1954 has an estimated distance measurement, $D= 1.67$ kpc \citep{Ortiz2013}, that is smaller than our distance determination for VVV-CL160.

\section{Possible association with NGC 6544}

{VVV-CL160} is located on the Galactic plane in the direction of the bulge, but it does not match the PMs of the disk nor of bulge field stellar populations; in fact, the differences are significant.
VVV-CL160 exhibits halo kinematics, with the PM vector pointing almost perpendicular to the Galactic plane. The only other known object that shares this motion is NGC~6544 (Fig. \ref{Figure2}); it is located only 5 deg away, which is relatively close in projection by Galactic standards.  

The massive GC NGC~6544 is located near the Galactic plane, at l=5.838 deg, b=-2.204 deg,  lies at a distance of $D=2.6$ kpc, and exhibits a  prograde orbit \citep{Vasiliev2021}. \citet{ContrerasRamos2017} and   \citet{Gran2021}  produced deep NIR CMDs for this GC, after properly selecting the PM member stars using VVV data. 
They measured $[Fe/H] = -1.44 \pm 0.04$ dex, $[\alpha /Fe] = 0.20 \pm 0.04$ dex, and $Age = 12 Gyr$ using the PARSEC isochrones \citep{Bressan2012, Marigo2017}.
This metallicity is consistent with the recent Apache Point Observatory Galactic Evolution Experiment survey (APOGEE) high-dispersion spectroscopic measurements of \citet{Nataf2019} and \citet{Mezaros2020}. The right panel of Fig. \ref{Figure2} shows 
%that the relative differences in color and magnitudes between these two clusters of similar metallicity are: $\Delta J-Ks = 0.13$ mag, and $\Delta Ks = 0.24$ mag. We also show
the isochrone fits for both clusters. For consistency with this previous work, we used the same alpha-enriched PARSEC isochrones for $[Fe/H]=-1.4$ dex and $Age=12$ Gyr, which give a good fit to our PM-selected VVV-CL160 members.

We verified the VVV-CL160 metallicity determination with the RGB slope in the NIR.
Using the calibration of \citet{Cohen2017}, for an observed slope=0.072, we obtain $[Fe/H]=-1.45 \pm 0.2$. This metallicity value is consistent with that of NGC 6544 and justifies our choice of isochrones (Fig. \ref{Figure2}).

VVV-CL160 has an unusually large PM for a Galactic GC, similar only to that of NGC 6544 (Fig. \ref{Figure4}), with which it may be associated according to its location, distance, and metallicity.
The total spatial separation between these two GCs, taking their difference in distance into account, would be about 1.5 kpc, which is also relatively close by Galactic standards.
NGC 6544 has a mean $PMRA = -2.30 \pm 0.03$ mas/yr, $PMDec = -18.60 \pm 0.03$ mas/yr \citep{Vasiliev2021}, similar to 
%. These are very similar to the mean PMs of 
VVV-CL160. In fact, the mean tangential velocity difference between these clusters is $\Delta V_t \sim 17$ km/s. Given their location in the Galaxy, this observed difference is largely accounted for as the Solar reflex motion implies a difference of $\Delta V_t \sim 15$ km/s and $\Delta RV \sim 20$ km/s.

It is quite unlikely to find two clusters so close together that share similar locations, metallicities, and extreme kinematics. We estimate that there is a $>99.99$ \% probability that these objects are associated.\footnote{
{As a first order estimate, the probability of finding two objects separated by only 5 degrees in the sky is 0.00785. The probability of finding two objects with the same extreme PMs out of a population of 150 objects is 0.0067. Since the two probabilities are independent, the total probability, P=1/20000, is simply obtained by multiplying these two quantities.}
} 
 This suggests that, even though they are too separated to be bound, they may still belong to the same structure.

The origin of NGC 6544 is not yet clear: It is not known if it is an accreted or an in  situ GC \citep{Horta2020}. NGC 6544 has recently been suggested to belong to very different structures according to different authors, including the Kraken \citep{Kruijssen2020}, the GES \citep{Bajkova2020}, the Koala \citep{Forbes2020}, the old halo \citep{ContrerasRamos2017, Riley2020}, the thick disk \citep{Perez-Villegas2020}, and an unassociated low-energy cluster \citep{Massari2019}. Below we give an alternative  interpretation for this GC, arguing that NGC~6544 may belong to an ancient dwarf galaxy that also comprises the Hrid halo stream and VVV-CL160.

\section{Possible link with the Hrid halo stream}

The Hrid halo stream was recently discovered by \citet{Ibata2020} as a thin stream located relatively nearby ($D \sim 3.1$ kpc).
Its member stars can be traced across 62 deg at intermediate Galactic latitudes ($10 < b < 25$ deg)  in the northern hemisphere before the contrast is lost and it disappears at low latitudes.
Follow-up spectroscopy of three member stars gives a mean metallicity of $[Fe/H]= -1.1$ \citep{Ibata2020}.

This new GC, VVV-CL160, is also located in the extension of the Hrid halo stellar stream, with consistent PMs, metallicities, and distances.
\citet{Ibata2020} computed the orbit of the Hrid stream, and the kinematics of both GCs, VVV-CL160 and NGC 6544
(PMRA and PMDec), are similar to those expected for the Hrid stream  extended into the Galactic bulge region (see their Fig. 8).
In addition, the distances are also consistent, but the radial velocity of VVV-CL160 still needs to be confirmed in order to fully trace its orbit. The radial velocity for NGC~6544 is $RV=-36.4$ km/s \citep{Mezaros2020}, different from both of the stars in VVV-CL160 measured by \citet{Borissova2014}. However, the differences do not significantly impact the orbital properties of VVV-CL160 (Fig. \ref{Figure5}).

\citet{Bonaca2021} specifically predicted that GCs associated with halo stellar streams might be far from their parent streams (tens of degrees away). Halo streams cannot be traced too close to the Galactic plane because of the severe field contamination. However, associating a GC located in the Galactic plane with a stellar stream allows us to project the full extension of the stream. In the case of  {VVV-CL160}, the possible association with the Hrid stream would prove that it is much more extended  than previously thought.
 The location of {VVV-CL160,} and also that of NGC~6544, matches the Hrid projected orbit and would extend this stream by $\sim 40$ deg more,
 %. This almost doubles the previously known extension, 
 making this one of the longest traceable halo streams.{
We stress the need for a more rigorous analysis when the radial velocities become available in order to confirm this association.}

We also have to consider the possibility that either VVV-CL160 or NGC~6544 may have been the nucleus of the Hrid parent dwarf galaxy. 
It would be interesting to measure their chemical inhomogeneities with follow-up high-dispersion spectroscopy in order to decide if they possess  multiple populations or a second generation of stars. 
% The relationship between globular cluster mass, metallicity, and light-element abundance variations needs to be explored for this new cluster. 
We argue that NGC~6544 is a better candidate to be the nucleus because it is more luminous and massive. There are recent element abundances measured for this cluster.  The sample analyzed by \citet{Nataf2019} only contained two stars for this cluster, and the spectroscopy of \citet{Mezaros2020} does not report multiple populations; however,  \citet{Gran2021}  find that about two-thirds of the APOGEE  spectra show distinct patterns in the individual element abundances and argue that they are possibly associated with a second generation of stars within the cluster.

\section{Conclusions}

We have discovered that VVV-CL160 is a large new low-luminosity metal-poor GC located in the Galactic plane in the direction of the bulge. We used VIRAC2 PMs to establish membership and measure the GC structural parameters. This GC is very concentrated, with a core radius of $r_c= 22.8"'$ (0.44 pc) and a tidal radius of $r_t = 50'$ (58 pc), and we estimate a total luminosity of $M_K=-7.6 \pm 0.3$ mag, equivalent to $M_V = -5.1$ mag.
Perhaps the most distinguishing property of this new GC is its peculiar PM, which is very different from that of the rest of the MW GCs -- with the exception of NGC 6544, which appears to have similar kinematics.

We made  NIR CMDs in order to estimate the cluster's main parameters via comparison with the well-studied Galactic GC NGC~6544.
We measured $E(J-Ks)= 1.95$ mag and $A_K = 1.40$ mag from the maps of \citet{Surot2020}, yielding a distance modulus of $m-M = 13.01 \pm 0.10$ mag, equivalent to $D=4.0 \pm 0.2$ kpc. We also estimated a mean metallicity of $[Fe/H]= -1.4 \pm 0.2$ dex from isochrone fits to the NIR CMD, adopting an age of $t = 12$ Gyr.
Desirable follow-up observations for this interesting new GC include (i) spectroscopy to measure its mean radial velocity and detailed chemical composition and (ii) deep imaging to measure its age using the main sequence turnoff.

{ Based on the location and kinematics, we suggest that there may be a connection between the Hrid stream, recently discovered in the northern hemisphere, and the GCs VVV-CL160 and NGC~6544, located in the southern hemisphere.\ If this is the case,  this stellar stream would be significantly longer than previously thought.
In fact, if the association of VVV-CL160 and NGC~6544 with the Hrid halo stream is confirmed with follow-up observations, this would upgrade Hrid from a mere stream to a fully fledged dwarf galaxy that is being accreted by the MW.
%like hte Sgr dwarf galaxy
%It would then not be surprising if the main body of the Hrid dwarf galaxy was located in the region of these two GCs, or
%It would also not be surprising 
%if there were 
It would then also be interesting to search for more GCs associated with this dwarf galaxy, which may be discovered at low latitudes, hidden behind high extinction. }

        \begin{acknowledgements}  
%       The authors are grateful for the enlightening feedback from the anonymous referee. 
      { We would like to thank the anonymous referee for the useful comments.}
      We gratefully acknowledge the use of data from the ESO Public Survey program IDs 179.B-2002 and 198.B-2004 taken with the VISTA telescope and data products from the Cambridge Astronomical Survey Unit. This work was developed in part at the Streams 21 meeting, virtually hosted at the Flatiron Institute. D.M. and M.G. are supported by Fondecyt Regular 1170121. D.M. is also supported by the BASAL Center for Astrophysics and Associated Technologies (CATA) through grant AFB 170002. This work has made use of data from the European Space Agency (ESA) mission \textit{Gaia} (\url{http://www.cosmos.esa.int/gaia}), processed by the \textit{Gaia} Data Processing and Analysis Consortium (DPAC, \url{http://www.cosmos.esa.int/web/gaia/dpac/consortium}). 
      %Funding for the DPAC has been provided by national institutions, in particular the institutions participating in the \textit{Gaia} Multilateral Agreement.      
\end{acknowledgements}
        
%\bibliographystyle{aa}
%\bibliography{references}

\begin{appendix}
\section{Comparison with the Hrid halo stream}

{
 Figures A1 and A2 show the comparison of the Hrid stream with the GCs VVV-CL160 and NGC 6544.
 These figures illustrate that the positions and PMs of these two GCs are consistent with an extension of the Hrid halo stream.
 This association, however, should be confirmed with future radial velocity measurements.
}
 
        \begin{figure}
        \begin{center}
                \includegraphics[width=82mm]{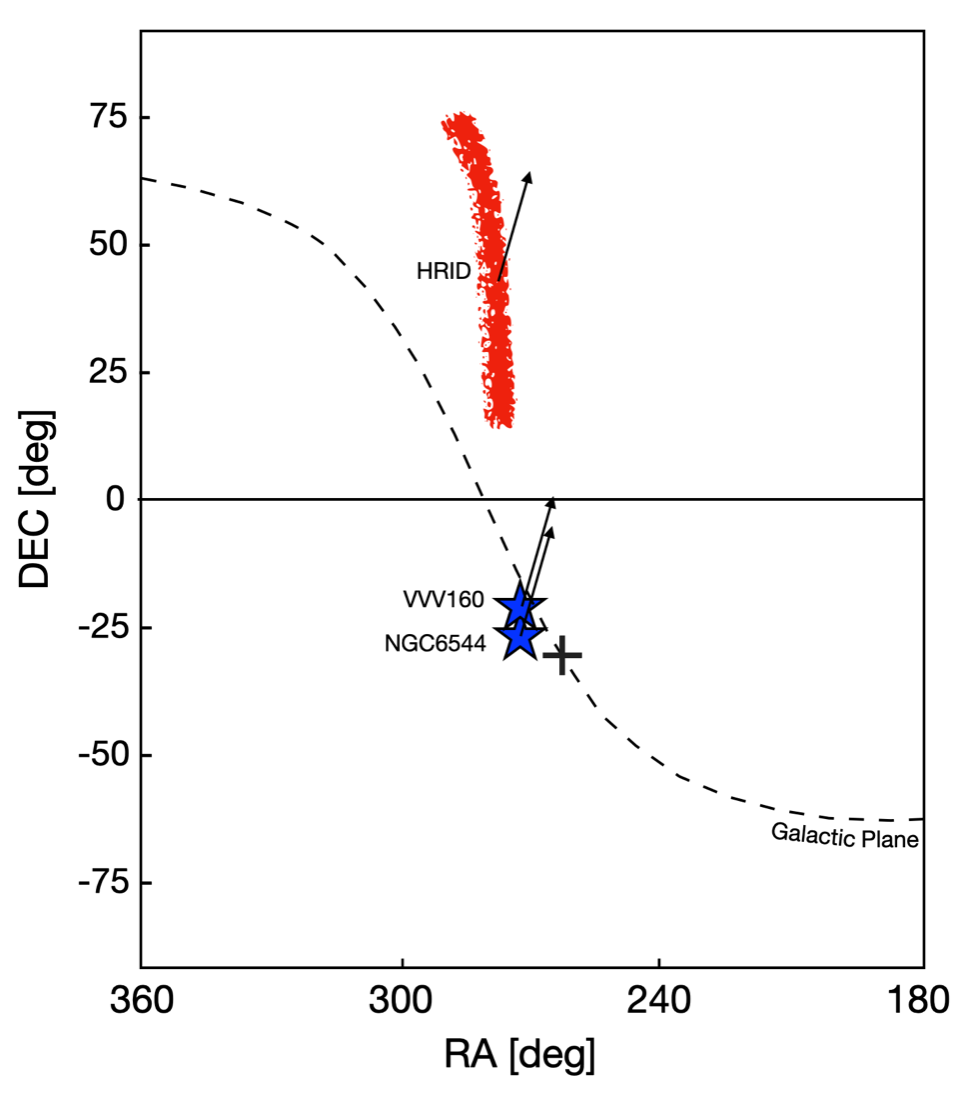}
                \caption{
{
                Position of the Hrid stream in equatorial coordinates compared with the GCs VVV-CL160 and NGC 6544.
                The arrows show the direction of the PMs of these objects, with the Hrid motion measured by Ibata et al. (2020).
                The Galactic plane is labeled, and the position of the Galactic center is shown with a cross.                           }
}
                \label{FigureA1}
        \end{center}
\end{figure}

        \begin{figure}
        \begin{center}
                \includegraphics[width=82mm]{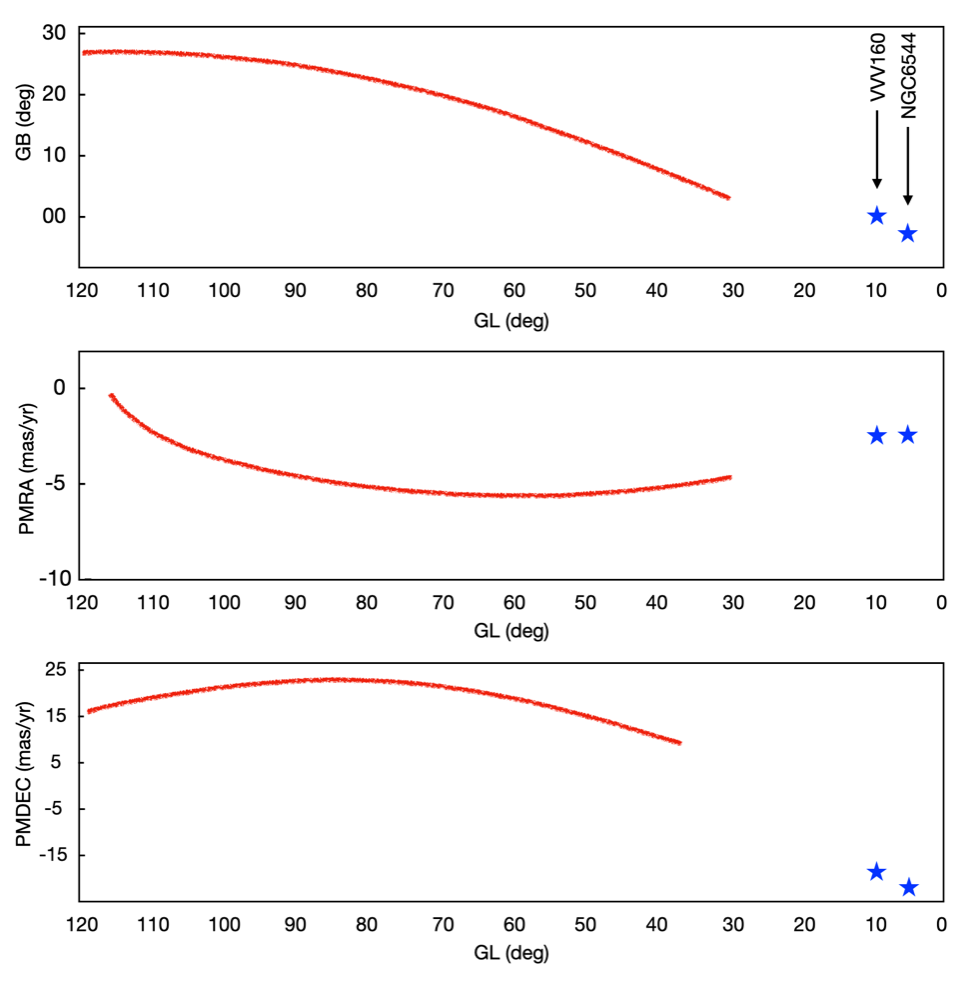}
                \caption{
{
                Comparison between the Hrid halo stream and the GCs VVV-CL160 and NGC 6544. 
                Top panel: Position of the Hrid stream in Galactic coordinates from Ibata et al. (2020) compared with the two GCs.
                Middle panel: PMRA for the Hrid stream as a function of Galactic longitude compared with both GCs.
                Bottom panel: PMDec for the Hrid stream as a function of Galactic longitude compared with both GCs.}
                                }
                \label{FigureA2}
        \end{center}
\end{figure}

\end{appendix}

\end{document}